\documentclass[AMA,STIX1COL]{WileyNJD-v2}
\usepackage{moreverb}

\newcommand\BibTeX{{\rmfamily B\kern-.05em \textsc{i\kern-.025em b}\kern-.08em
T\kern-.1667em\lower.7ex\hbox{E}\kern-.125emX}}

\articletype{Research Article}%

\received{<day> <Month>, <year>}
\revised{<day> <Month>, <year>}
\accepted{<day> <Month>, <year>}

\newcommand{\sapphire}{sapphire~}
\newcommand{\sapphireNoSpace}{sapphire}
\newcommand{\alscn}{Al\textsubscript{0.74}Sc\textsubscript{0.26}N~}
\newcommand{\alscnNospace}{Al\textsubscript{0.74}Sc\textsubscript{0.26}N}
\newcommand{\alscnX}{Al\textsubscript{1-x}Sc\textsubscript{x}N~}
\newcommand{\AlBN}{Al\textsubscript{1-x}B\textsubscript{x}N}
\newcommand{\alscnXNospace}{Al\textsubscript{1-x}Sc\textsubscript{x}N}
\newcommand{\substratePoly}{Pt/Ti/SiO\textsubscript{2}/Si}
\newcommand{\substrateEpi}{Pt/GaN/sapphire}

\begin{document}

\title{In-Grain Ferroelectric Switching in Sub-5 nm Thin \alscn Films at 1 V}

\author[]{Georg Schönweger$^{1,3,\partial}$, Niklas Wolff$^{2,4,\partial}$, Md Redwanul Islam$^{2}$, Maike Gremmel$^{2}$, Adrian Petraru$^{1}$, Lorenz Kienle$^{2,4}$, Hermann Kohlstedt$^{1,4}$, Simon Fichtner$^{2,3}$}

\address[1]{\orgdiv{Department of Electrical and Information Engineering}, \orgname{Kiel University}, \orgaddress{Kaiserstrasse 2, D-24143 Kiel, Germany}}
\address[2]{\orgdiv{Department of  Material Science}, \orgname{Kiel University}, \orgaddress{Kaiserstrasse 2, D-24143 Kiel, Germany}}
\address[2]{\orgdiv{Fraunhofer Institute for Silicon Technology (ISIT)}, \orgaddress{Fraunhoferstr. 1, D-25524 Itzehoe, Germany}}
\address[4]{\orgdiv{Kiel Nano, Surface and Interface Science (KiNSIS)}, \orgaddress{Kiel University, Christian-Albrechts-Platz 4, D-24118 Kiel, Germany}}

\corres{*Georg Schönweger, \email{gmsc@tf.uni-kiel.de}, Simon Fichtner, \email{sif@tf.uni-kiel.de}
\\
$^\partial$ Authors contributed equally to this work.}

\abstract[Abstract]{
Analog switching in ferroelectric devices promises neuromorphic computing with highest energy efficiency, if limited device scalability can be overcome. To contribute to a solution, we report on the ferroelectric switching characteristics of sub-5 nm thin \alscn films grown on \substratePoly~ and epitaxial \substrateEpi~templates by sputter-deposition. In this context, we focus on the following major achievements compared to previously available wurtzite-type ferroelectrics: 1) Record low switching voltages down to 1~V are achieved, which is in a range that can be supplied by standard on-chip voltage sources. 2) Compared to the previously investigated deposition of thinnest \alscnX films on epitaxial templates, a significantly larger coercive field to breakdown field ratio is observed for \alscn films grown on silicon substrates, the technologically most relevant substrate-type. 3) The formation of true ferroelectric domains in wurtzite-type materials is for the first time demonstrated on the atomic scale by scanning transmission electron microscopy investigations of a sub-5 nm thin partially switched film. The direct observation of inversion domain boundaries within single nm-sized grains supports the theory of a gradual domain-wall motion limited switching process in wurtzite-type ferroelectrics. Ultimately, this should enable the analog switching necessary for mimicking neuromorphic concepts also in highly scaled devices.}


\maketitle

\section{Introduction}
In recent years, ferroelectrics have become one of the main foci of advancing semiconductor technology towards higher performance and energy efficiency.\cite{Khan2020, Mikolajick2020,Mikolajick2021} This applies especially to neuromorphic and in-memory computing, where the field-driven ferroelectric effect promises analog operation with the lowest input power.
However, the entrance of ferroelectric functionality into the active areas of commercial devices other than binary ferroelectric random-access memories (FRAMs) is yet to take place. One of the major challenges in this context is an excess of device-to-device variability of key parameters like the threshold voltage in small devices as well as the loss of their capability to operate in an analog fashion. This variability becomes pronounced when the ferroelectrically active area of a device approaches the size of the grains inside the ferroelectric films and the domains therein. This grain size typically is in the range of tens of nanometer for the fluorite-type ferroelectrics, which have been at the focus of scientific attention in recent years.\cite{Schroeder2022}
The factors contributing to device variability are a lack of crystalline texture, stress inhomogeneities and less than complete phase purity, which lead to different material parameters between different grains.\cite{Schroeder2022,Lederer2019} The possibility of analog operation in turn becomes compromised due to the nucleation limited switching of the fluorite-type films. This implies that ferroelectric domains quickly reach their final shape while nucleating, leading to a digital behavior in devices where only a small number of ferroelectric domains remain inside the active area.\cite{Mulaosmanovic2017}

Since their discovery in 2019, the new wurtzite-type ferroelectrics have raised expectations of a possible solution to the aforementioned issues.\cite{Fichtner2019} Wurtzite-type ferroelectric films can typically be grown phase pure, well textured and the narrow distribution of their displacement current response upon ferroelectric switching promises a narrow distribution of the local ferroelectric properties, all of which should contribute to improved device repeatability. At the same time, wurtzite-type films can easily be deposited at complementary metal oxide semiconductor (CMOS) back-end-of-line (BEOL) compatible conditions, feature extreme temperature stability themselves and thicker films are already in large-volume industrial production.\cite{Islam2021,Aigner2019} Further, for film thicknesses above 100 nm, the switching kinetics of the material can be modeled to be domain wall motion limited,\cite{Fichtner2020} which implies a gradual or analog switching on the atomic level. Despite these conceptual advantages, major challenges remain to be solved until the material class is able to fully meet the demands of advanced microelectronic devices. In this context, it is highly necessary to further reduce the ferroelectric switching voltage of wurtzite-type thin films to meet the capabilities of typical on-chip voltage supplies (in the range of 1 V), while retaining the aforementioned advantages like phase purity and domain wall motion limited switching.

In this study we demonstrate for the first time that wurtzite-type sub-5 nm thin (8 to 9 unit cells corresponding to 4 - 4.5 nm) \alscn films sputter-deposited on silicon (Si) substrates retain ferroelectric functionality with switching voltages as low as 1 V and feature in-grain, nm-sized domains upon partial switching. We were thus able to reduce the switching voltage and film thickness of films on Si by around 50\% compared to literature (5 nm thin films grown by non-BEOL compatible molecular beam epitaxy (MBE) and $\approx 10$ nm thin films grown by sputtering on Si).\cite{Wang2023,Mizutani2021,Yasuoka2022} Further, by performing atomic resolution scanning transmission electron microscopy (STEM) on epitaxial sub-5 nm thin \alscn films, we obtained the first images of domain walls in any wurtzite-type ferroelectric to confirm the presence of nm-sized domains within individual grains. 

Our investigation starts with a structural as well as an electrical comparison of 10 nm thin \alscn grown epitaxially on \substrateEpi~and grown non-epitaxially on \substratePoly~to demonstrate the improved ferroelectric properties of the latter. Further downscaling to the sub-5 nm range of ferroelectric \alscn films grown on Si is investigated. The scaling of the coercive voltage, including a decrease of $E_c$ below 10 nm film thickness ultimately allowed to achieve switching voltages as low as 1 V and is discussed in detail. Epitaxial growth was investigated as well, as it allows to resolve the ferroelectric polarization reversal on atomic level in partially switched sub-5 nm thin \alscn layers via STEM. The identification of regions with opposite polarity inside a single grain and the necessary occurrence of a domain boundary in between gives first insights into the size, shape, location and evolution of ferroelectric domains in \alscn and potentially in the whole class of wurtzite-type ferroelectrics.

\section{Results $\&$ Discussions}
\subsection{Effect of non-epitaxial growth on Si vs. epitaxial growth on GaN on the ferroelectric response of \alscn}
For the direct integration of ferroelectric wurtzite-type films to CMOS technology, the possibility to deposit them on Si substrates without epitaxial templating is crucial. While one might assume that epitaxial growth and thus higher interface- and film quality will automatically result in improved electrical properties, this section motivates that the opposite can be the case for \alscnXNospace.

\begin{figure}[h!]
\centering
\includegraphics[]{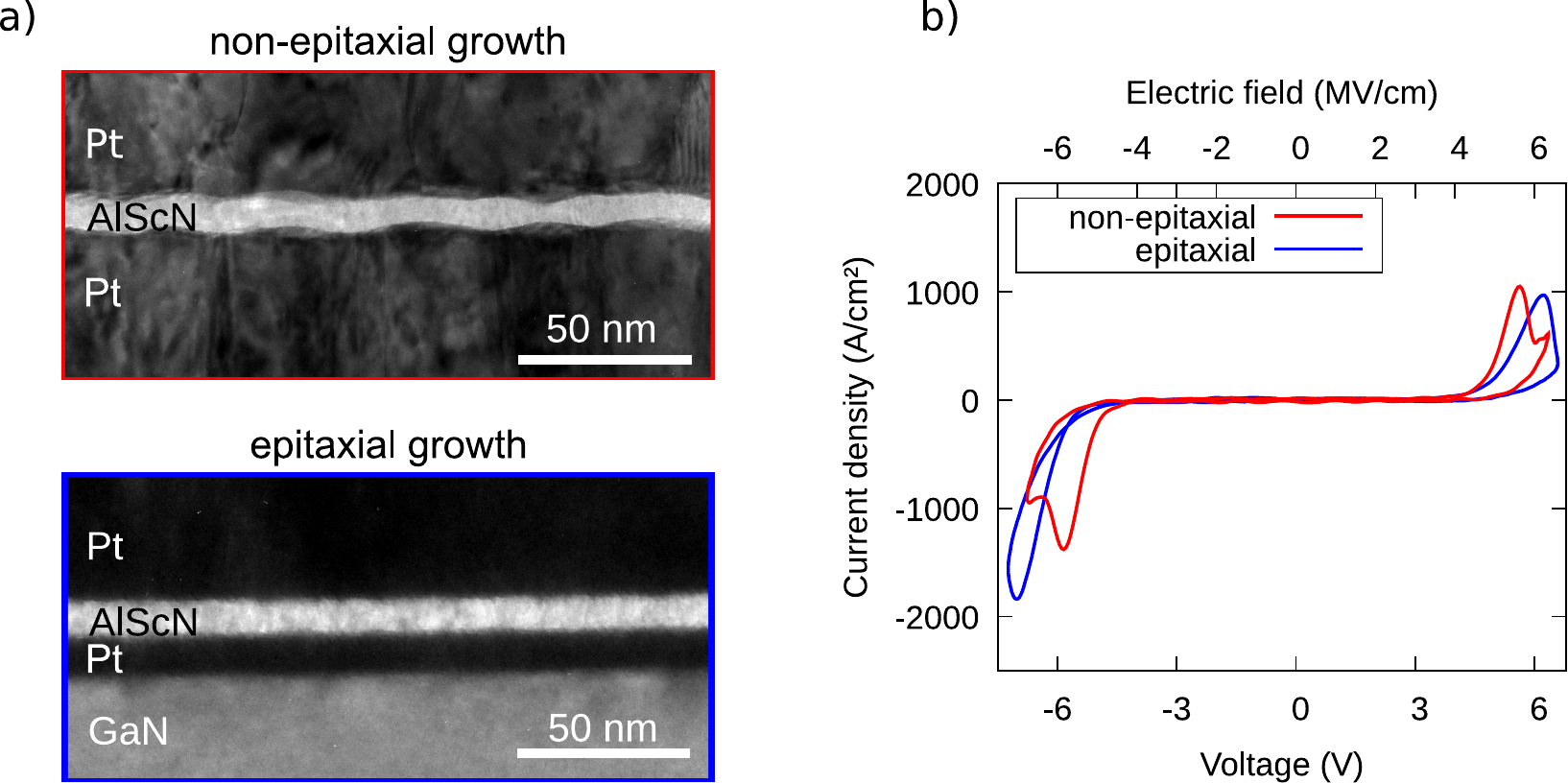}
\caption{a) TEM cross-section of 10 nm thin \alscn grown (top) non-epitaxially within a Pt/\alscnNospace/\substratePoly~and (bottom) epitaxially within a Pt/\alscnNospace/Pt/GaN/\sapphireNoSpace~capacitor stack. Only the capacitor structures are depicted. b) The $J-E$ loops of the capacitors depicted in a), measured at 100 kHz.}
\label{fig:tem-iv-10nm}
\end{figure}

This can be concluded from the electrical response as well as from the interface quality investigations of 10 nm thin \alscn films. In Figure \ref{fig:tem-iv-10nm}a, the cross-sections of 10 nm thin \alscn films grown epitaxially on \substrateEpi~ as well as grown non-epitaxially on \substratePoly~ are compared. All films were capped $in~situ$ to prevent oxidation of the \alscn surface, which is crucial to obtain an undisturbed ferroelectric response especially of < 10 nm thin films, where the thickness of the native oxide can be in the range of the total film thickness.\cite{Schoenweger2022a,Wang2020,Li2022} 

For the epitaxially grown film stacks, the interfaces are smooth with an overall low surface roughness of the respective layers, which is known to result in a reduction of the leakage currents in capacitors.\cite{Zhao1999} Structurally, the epitaxial films with a 10 nm thin Pt bottom electrode layer also have a superior crystalline quality compared to non-epitaxial ones, i.e., higher c-axis texture, which we investigate in detail in a separate work.\cite{redwan2023} Nonetheless, the films deposited on Si substrates exhibit more pronounced ferroelectric switching peaks (Figure \ref{fig:tem-iv-10nm}b). Thus, despite their higher interface roughness and poorer interface texture compared to the epitaxial ones, a complete polarization inversion is demonstrated for the non-epitaxial 10 nm thin \alscn grown on \substratePoly, yet not for the epitaxial films. This is apparent from the drop in current density after ferroelectric switching at the coercive field ($E_c$) with maximum $J$ followed by a local minimum before the contribution from leakage currents leads to a further increase in $J$. In comparison, no local minimum is observed for the epitaxial film. Although the leakage currents for the 10 nm thin epitaxial \alscn grown on \substrateEpi~ are lower (at a fixed voltage) compared to films grown on Si, $E_c$ is also higher and approaches the electrical breakdown field. Thus, as demonstrated in our recent work, we were able to fully switch the polarization of 10 nm thin epitaxial films via capacitance vs. electric field $C-E$ measurements, but not via $J-E$ loops.\cite{Schoenweger2022a} We attribute the improved $E_c$ of the films grown on Si compared to the ones grown on \sapphire to differences in the respective \alscn film stress, which is well known to result in a shift of $E_c$.\cite{Fichtner2019} Despite the fact that both heterostructures were grown under exactly the same \alscn deposition conditions (same run), the thermal expansion coefficients ($\alpha_{sub}$) of the silicon substrate (non-epitaxial growth - $\alpha_{sub} = 2.6 x 10^{-6}$/K) and \sapphire substrate (epitaxial growth - $\alpha_{sub} = 7.3 x 10^{-6}$/K) differ, leading to strong differences in the thermally induced film stress after cooling down from the \alscn ($\alpha_{sub} = 4.9 x 10^{-6}$/K) deposition temperature at 450 °C.\cite{Okada1984,Yim1974,Lu2018} Thus, in addition to the film-stress induced by interface strain, grain boundaries and defects, tensile stress is thermally induced in \alscnX if grown on a silicon substrate, while compressive stress is thermally induced if grown on a \sapphire substrate.

The ability to tune the coercive field of \alscnX exploiting the thermal expansion of varying substrates has also been reported recently by Yasuoka et al.\cite{Yasuoka2022a} In consequence, thermal induced tensile stress extends the in-plane lattice resulting in the reduction of $E_c$, similar to an increase in Sc concentration. To conclude, we attribute the more pronounced ferroelectric displacement current peak of the non-epitaxial 10 nm thin \alscn to a more favorable position of $E_c$ compared to the onset of leakage (compare the local minima in the current response) and with respect to the breakdown strength. 

\subsection{Ferroelectric properties of sub-5 nm thin \alscn films grown on Si}

Next, we present and discuss the electric characterization results of sputter-deposited 8 - 9 unit cells (4 - 4.5 nm) thin \alscn films grown on Si wafers. Details on the exact thickness determination by using STEM can be found in section \ref{sec4}.

\begin{figure}[h!]
\centering
\includegraphics[]{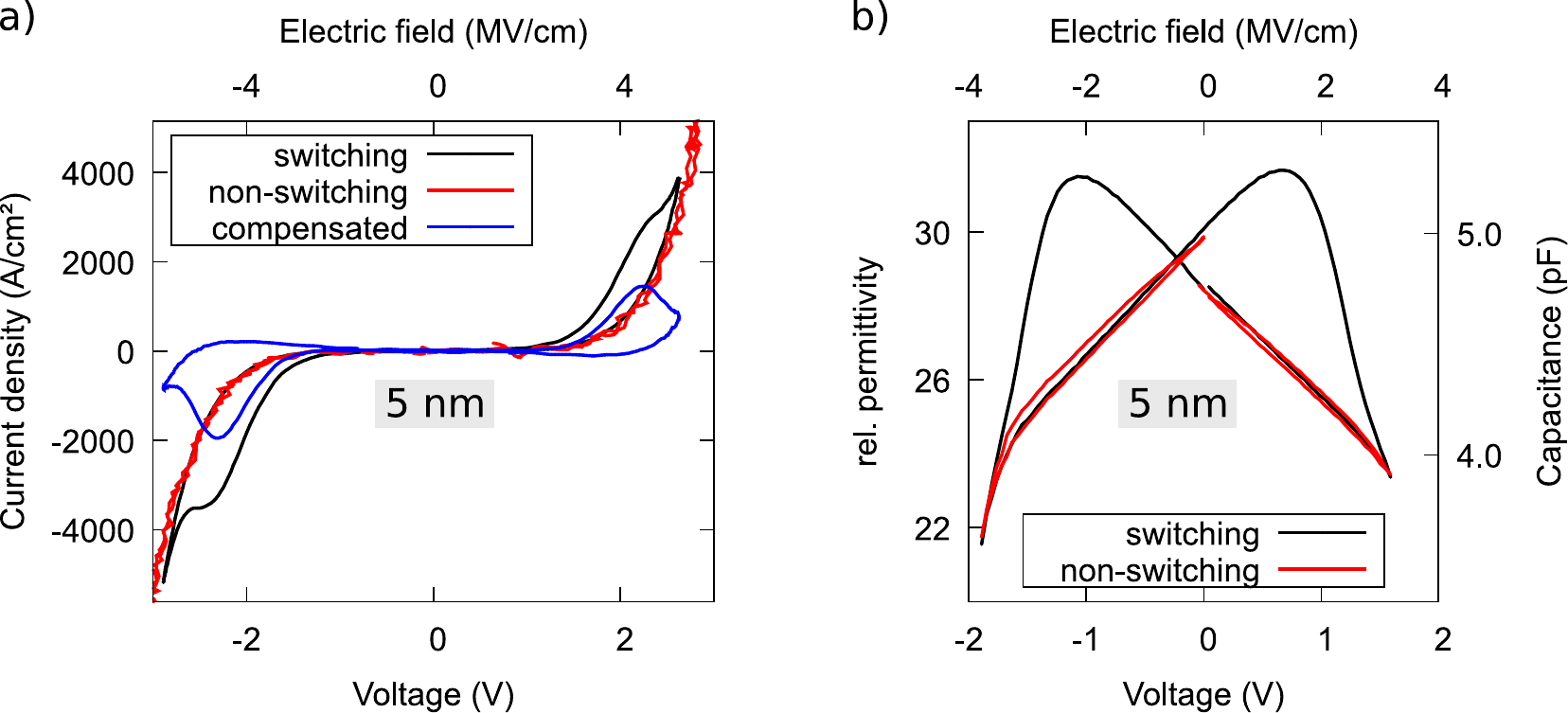}
\caption{a) $J-E$ loops of sub-5 nm thin \alscn grown on \substratePoly~measured at 100 kHz on 5 \textmu m diameter pads. A leakage current compensated curve (blue) by subtracting the non-switching- (red) from the switching currents (black) is included. b) $C-V$ loop of the sub-5 nm thin \alscn based capacitor described in a) measured on a 10 \textmu m diameter pad. Unipolar non-switching cycles (red) by measuring each branch (positive and negative voltages) twice with the same polarity are included to stress the non-volatile nature of the permittivity enhancement due to ferroelectricity.}
\label{fig:I-V-Pt-Si-10To5nm+C-V-5nm}
\end{figure}

In Figure \ref{fig:I-V-Pt-Si-10To5nm+C-V-5nm}a, $J-E$ loops of sub-5 nm thin \alscn grown on \substratePoly~are illustrated. In direct measurements (black curve), the clear hysteresis is already indicative of ferroelectric switching. The leakage current flow through the dielectric as well as the displacement current contributions due to the relative permittivity ($\epsilon_r$) can be separated from the hysteretic (i.e., ferroelectric) displacement currents by recording non-switching loops (i.e., by pre-poling the respective measured positive and negative branch). After substraction of the non-switching- (red curve) from the switching currents (black curve) the typical shape of ferroelectric displacement current peaks are obtained (blue curve), which allows the extraction of $E_c$.

The $C-V$ loop of the sub-5 nm thin \alscn based capacitor depicted in Figure \ref{fig:I-V-Pt-Si-10To5nm+C-V-5nm}b further confirms the ferroelectric nature of the hysteretic event. A distinct butterfly-shaped loop, typical for ferroelectric switching, is visible. Clearly distinguishable, non-hysteretic non-switching loops are depicted as well for the $C-V$ loop. Furthermore, the polarization inversion of a sub-5 nm thin film is unambiguously demonstrated by atomically resolved STEM investigations discussed in section \ref{sec4}. Thus, it is demonstrated that such thin ferroelectric wurtzite-type films can be grown by sputter deposition on oxidized silicon in a manner compatible with CMOS technology, which is a clear advantage over high-temperature ($\geq 500$ \textdegree{}C) MBE deposition processes on single crystal templates. \cite{Wang2023, Wang2020}

Furthermore, the ferroelectric switching of \alscn films grown on silicon with, for wurtzite-type materials, record low voltages down to 1 V is a major milestone towards ferroelectric \alscnX based future devices operable with the on-chip voltage supply of integrated circuits.\cite{Maxfield2008,Chen2009}

\subsection{Coercive field scaling in ultrathin \alscn}

The low switching voltage down to 1 V reached in our films is not only due to a simple reduction in thickness, but also due to the favorable scaling of $E_c$ with thickness, which we will therefore discuss in more detail in this section. In particular, the appearance of a depolarization field and its effects on the electrical response of films below 10 nm film thickness are discussed, as is the relative dielectric permittivity - which itself is related to the coercive field through the shape of the ionic potential wells.\cite{Mikolajick2021}

\begin{figure}[h!]
\centering
\includegraphics[]{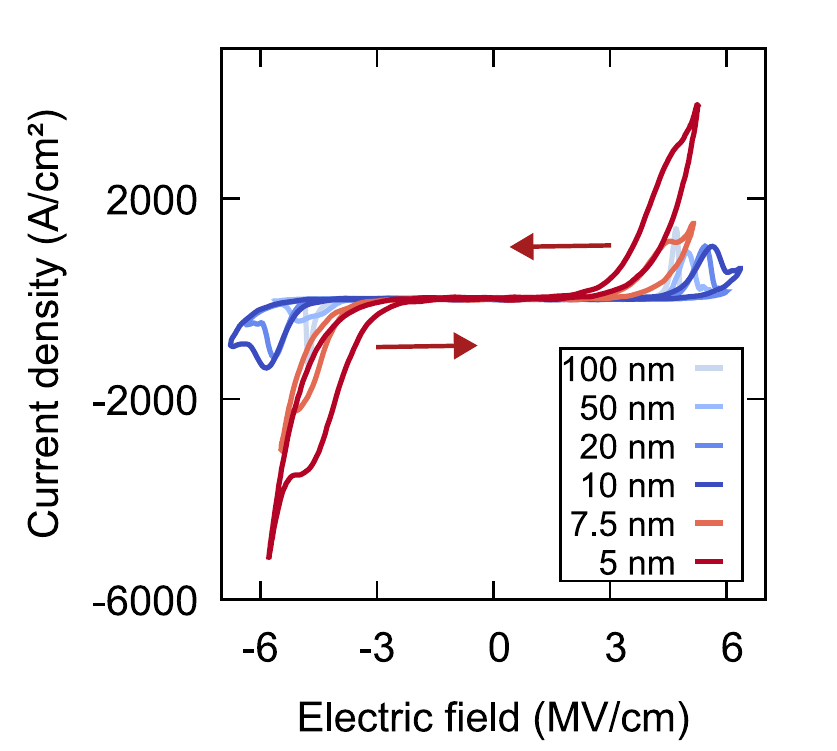}
\caption{$J-E$ loops of 100 nm- down to sub-5 nm thin \alscn based capacitors deposited on \substratePoly. All measurements were performed at 100 kHz on 5 \textmu m diameter pads (< 10 nm \alscn thickness) and 10 x 10 \textmu m² pads (> 10 nm \alscn thickness). The decreasing trend of $E_c$ with decreasing film thickness below 10 nm is indicated by red arrows.}
\label{fig:I-V-Pt-Si-100nm-To-5nm}
\end{figure}

In Figure \ref{fig:I-V-Pt-Si-100nm-To-5nm}, $J-E$ loops of 100 nm- down to sub-5 nm thin \alscn based capacitors are depicted. From 100 nm down to 10 nm the coercive field is increasing with decreasing film thickness, but interestingly, below 10 nm the coercive field is significantly decreasing again, as indicated by the red arrows in Figure \ref{fig:I-V-Pt-Si-100nm-To-5nm}. A comparable trend with thickness scaling down to sub-5 nm is also observed for epitaxial films grown on \substrateEpi~ (see Supplement - Figure \ref{fig:Supp.Ec-vs-thickness-epiPt}), hence it is concluded that the scaling properties are rather independent of the substrate (silicon vs. \sapphireNoSpace), crystalline quality and associated growth modes (non-epitaxial vs. epitaxial).

\begin{figure}[h!]
\centering
\includegraphics[]{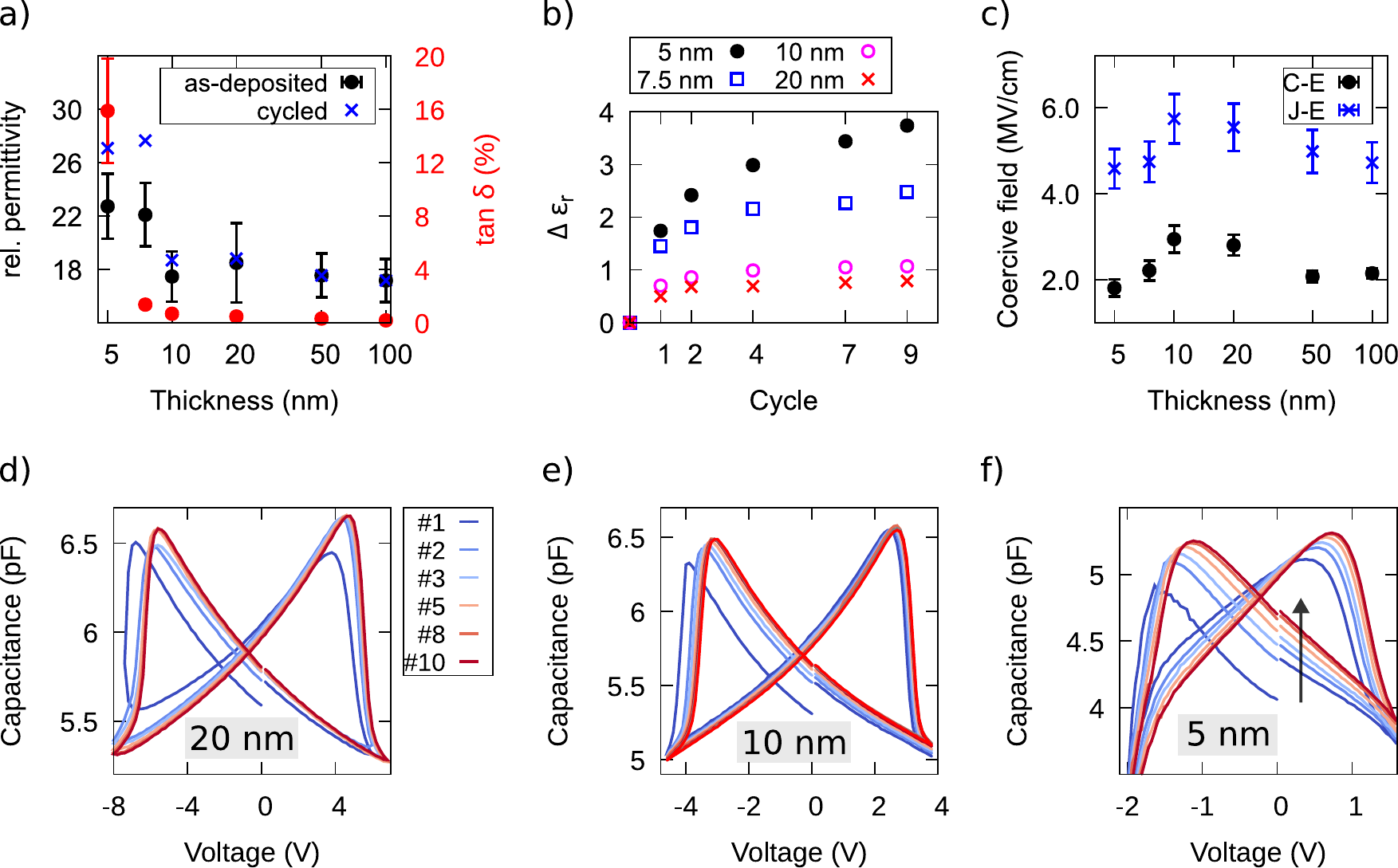}
\caption{a) Relative permittivity as well as loss tangent as a function of \alscn film thickness for as-deposited- and for pre-cycled (10 times) capacitors grown on \substratePoly. b) Absolute change of $\epsilon_r$ (state with full positive polarization, 0 V bias) with cycling in dependence of the \alscn film thickness. c) The coercive field dependence on \alscn thickness for the capacitors described in a) determined via $J-E$ (100 kHz) and via $C-E$ (sweep time 20 s, small signal 100 mV and 900 kHz) loops. The coercive field determined via $C-E$ loops is approximated by the peak positions of the butterfly-loop. d) First ten $C-V$ cycles of pristine capacitors consisting of 20 nm thin-, e) 10 nm thin- and f) sub-5 nm thin \alscn used for determining the change of $\epsilon_r$ with cycling as depicted in b). The capacitor area was 695 \textmu m² (20 nm thickness), 341 \textmu m² (10 nm thickness) and 99 \textmu m² (sub-5 nm thickness).}
\label{fig:e_r.A.tanD-++Ec-vs-thickness}
\end{figure}

A slight increase in $E_c$ with decreasing film thickness down to 10 nm is consistent with the scaling properties reported so far for \alscnXNospace.\cite{Schoenweger2022a, Yasuoka2022,Mizutani2021, Wang2023} Yasuoka et. al attributed this behavior to a change in the lattice-parameters for thinner films due to stress gradients arising from the lattice mismatch between Pt (2.78 \r{A}) and \alscnX (3.22 \r{A} for  $x=0.2$). Despite the high lattice mismatch, an epitaxial-like growth between Pt grains of the bottom-electrode layer and \alscnX grains is suggested, eventually resulting in an increase in compressive strain in the basal plane when reducing the film thickness. In our films, the lattice-parameters do not change significantly for thicknesses down to 10 nm. However, for sub-5 nm thickness the \alscn lattice-parameters determined via STEM are $a \approx 319$ pm and $c \approx 505$ pm (details on the determination can be found in the Experimental section). This implies (relative to the equilibrium $a$-lattice parameter of $\approx 324$ pm at a Sc concentration of $x = 0.26$) an in-plane compressive strain of $\approx 1.5$\%.\cite{Ambacher2021a,Schoenweger2022}
However, we measure $E_c$ to decrease below 10 nm film thickness down to less than 2 MV/cm in $C-E$ curves, as illustrated in Fig. \ref{fig:e_r.A.tanD-++Ec-vs-thickness}c. This decrease in $E_c$ below 10 nm thickness differs from the very recently reported thickness scaling study down to 5 nm thin epitaxial films grown via MBE.\cite{Wang2023} In this work, Wang et al. also attributed the increase in $E_c$ to a stress gradient forming due to the smaller in-plane lattice-parameter of the Mo bottom electrode compared to the one of \alscnXNospace.

The reduction in the electric field necessary for switching in films thinner than 10 nm is especially pronounced when considering the onset of the hysteresis opening in the $J-E$ loops, as visible in Figure \ref{fig:I-V-Pt-Si-100nm-To-5nm}. For sub-5 nm film thickness the hysteresis opens at 2.1 MV/cm, while for 100 nm film thickness, the opening starts at 4.3 MV/cm. This implies a more gradual switching capability below 10 nm film thickness.

 A decrease in $E_c$ in ultrathin ferroelectrics was reported by Dawber et al., who included depolarization field corrections into the Janovec-Kay-Dunn scaling.\cite{Dawber2003,Kay1962,Janovec1958} The depolarization field ($E_d$) resulting from a finite screening length in the electrodes adds up to the applied electric field if the condition $4 \pi P_s >> \epsilon_r \epsilon_0 E$ is fulfilled. For \alscnNospace, with a spontaneous polarization ($P_s$) of $\approx 110$ \textmu C/cm² and electric fields ($E$) up to 6 MV/cm, this condition is clearly satisfied ($1507 >> 14$). Thus, similar to what was experimentally observed in ferroelectric PVDF films, a thickness dependent depolarization field qualitatively fits very well to the drop of $E_c$ below 10 nm film thickness in \alscnXNospace.\cite{Ducharme2000} Very recently, it has also been demonstrated by first-principles calculations that reducing the thickness ($d$) of usually non-switchable Wurtzite III-V semiconductors (e.g., AlSb) could result in polarization switching capability (i.e., ferroelectricity), due to the depolarization field which scales with $\varpropto 1/d$.\cite{Ke2023} 
 
The reason why Wang et al. did observe an increasing $E_c$ down to 5 nm is most likely due to the small remanent polarization ($P_r$) of their films. A $P_r$ of around 20~\textmu C/cm$^2$ at the given composition cannot itself be a consequence of solely the depolarization field without being accompanied by severe retention issues - which the group did not observe in their recent paper.\cite{Wang2023} An intrinsically lower $P_r$ however will lead to a proportionally reduced depolarization field.\cite{Black1997} Thus, the increased compressive stress in thinner films leading to a higher $E_c$, is not compensated to the same degree as in our work.
 
 The decrease in $E_c$ in our work is also reflected in the increase in $\epsilon_r$ below 10 nm film thickness, as illustrated in Figure \ref{fig:e_r.A.tanD-++Ec-vs-thickness}a. In addition, $\epsilon_r$ increases to even higher values after cycling, which is especially pronounced for the thinner films (Figure \ref{fig:e_r.A.tanD-++Ec-vs-thickness}b). A similar increase in the relative permittivity with cycling has also been observed for the wurtzite-type ferroelectric \AlBN.\cite{albn_wakeup} Through analysis of the Rayleigh parameters, this increase has been related to an increase in domain wall area compared to pristine samples at 0 V bias. 
If persistent domain walls indeed form during cycling and these domain walls extend vertically in the film, similar to what is reported in the following section, an enhancement in permittivity with lower film thickness would be a natural consequence - due to an increase in the ratio of domain wall area to film volume with reduced thickness. This change in the ratio would imply a larger relative volume that is frustrated by the domain wall and in turn features a higher permittivity due to shallower ionic potential.

With decreasing film thickness not only the leakage current-, but also the hysteretic area is increasing especially for sub-5 nm film thickness, as illustrated in Figure \ref{fig:I-V-Pt-Si-100nm-To-5nm}. This increase in apparent displacement current can not be explained alone by polarization reversal, as it would imply a physically unlikely large spontaneous polarization in excess of 1000 µC/cm². The enhanced apparent polarization has therefore to be attributed to a dynamic current contribution triggered by the polarization reversal of the \alscn film. Currently, the most likely explanation of this behavior is the temporally formation of conductive domain walls during switching.\cite{EvansGarciaMeierBibes+2020} As discussed above, with decreasing film thickness, the relative domain-wall density will increase and domains are more likely to extend from the top to the bottom interface. Both effects can facilitate increased electrical current to flow in the form of compensation charges for the strong polarization discontinuity along the domain walls. This concept of conducting domains walls is closely related to the well known polarization doping schemes in III-N semiconductors.\cite{Jena2002,Yan2018,Khokhlev2013} Further analysis of this effect is the focus of ongoing work.

\subsection{Atomic scale investigation of ferroelectric domains in sub-5 nm thin \alscnNospace} \label{sec4}

Analog-like ferroelectric switching is an elegant approach for emulating synapses and thus a stable partially switched state is an essential material property in the context of neuromorphic computing.\cite{Islam2019,Shi2021} Hence, it is important to image and understand the atomistic switching processes and the evolution of polarization discontinuities (i.e., ferroelectric domain walls). This section explores the microscopic consequences of ferroelectric switching in sub-5 nm thin \alscn films as well as their general structural properties via high-resolution STEM. The main focus of this study is the first observation of domain walls in individual \alscn grains in any wurtzite-type ferroelectric. In order to clearly observe the local polarization on unit cell scale, the analysis was conducted on an epitaxial (in-plane ordered), yet still polycrystalline film (0002 oriented columnar grains), which allows for direct imaging conditions because of the identical film/substrate crystallographic orientation. An overview image of the sub-5 nm thin \alscn film showing individual epitaxial grains with c-axis texture confirmed across the entire prepared area as well as $C-E$ loops demonstrating the ferroelectric switching in 10 nm- down to sub-5 nm thin epitaxially grown films is provided in the Supplement (Figure \ref{fig:FigS_texture} and Figure \ref{fig:C-V_GaN_thickness}). While previous attempts to resolve the local polarization in wurtzite-type ferroelectrics were successful to identify the polarization direction of single unit cells, the observation of domain walls has so far been elusive. 

The epitaxial nature of the heterostructure allowed for an accurate thickness determination on the level of monolayers due to the atomically sharp interfaces (c.f., Figure \ref{fig:tem-iv-10nm}a). Despite the columnar growth mode of sputtered films, the good structural quality of the in-plane oriented growth enables the direct observation of the unit-cell polarity within single grains on the atomic scale.\cite{Wolff2021} In order to draw conclusions on the switching process itself (besides just confirming up and down polarization flips), the investigated capacitor was only partially switched from the Nitrogen (N)-polar to the metal (M)-polar state. For this, the capacitor was pre-switched to full N-polarity by applying a positive field which is high enough to saturate the polarization reversal, with subsequent application of a negative field which is $\approx 0.8$ MV/cm below the saturation point.

\begin{figure}[h!]
\centering
\includegraphics[]{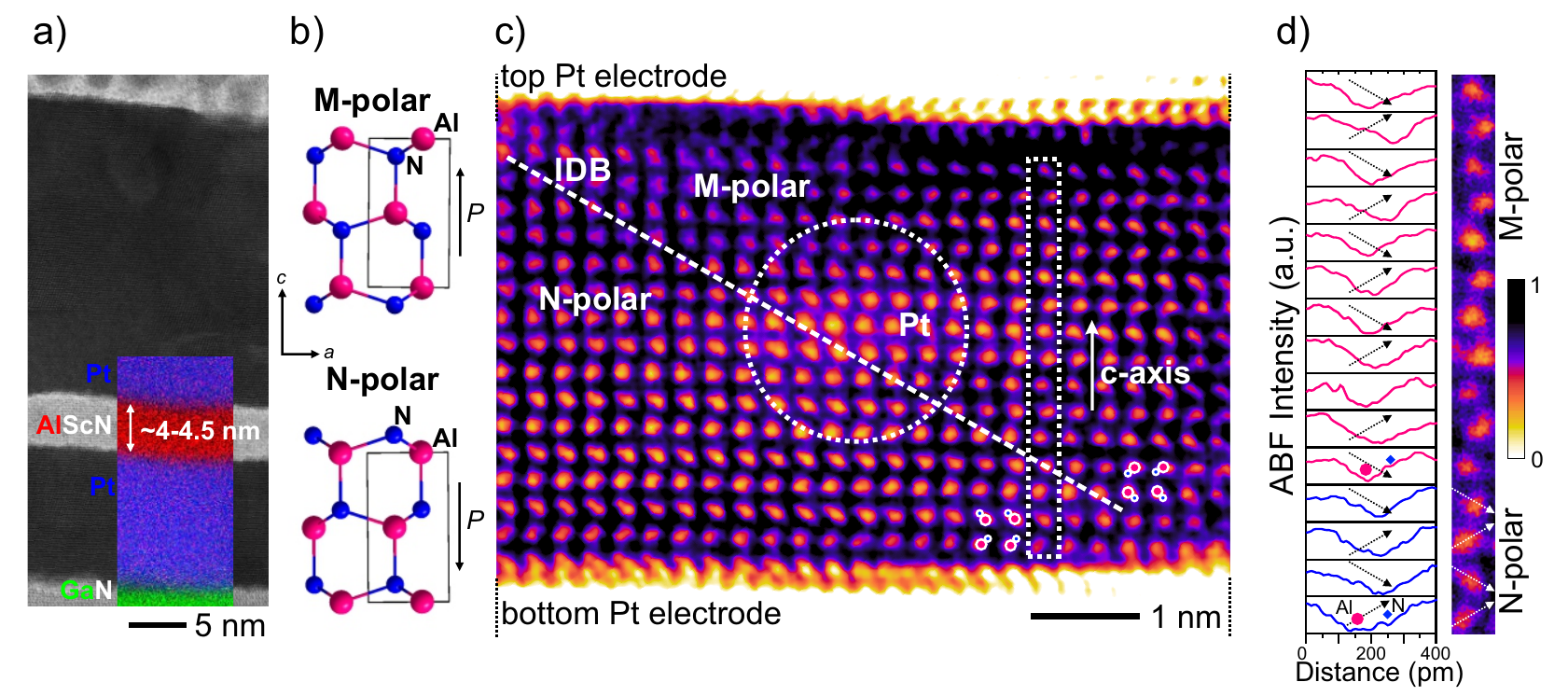}
\caption{a) ABF-STEM micrograph showing the Pt/\alscnNospace/Pt/GaN capacitor stack in cross-section. The inset shows the superimposed EDS maps of Pt, Al and Ga. b) Sketches of the atomic structure in the M- and N-polar state along the line of sight. c) Inverted-ABF-STEM micrograph of the full \alscn layer featuring an inclined inversion domain boundary separating regions of M-polarity (upper right) and N-polarity (lower left). Superimposed sketches of the (Al,Sc)-N dumbbells help to visualize the polarization direction. d) Intensity profile analysis of the polarization direction of individual (Al,Sc)-N dumbbells inside the single-column frame. Profiles are always drawn from left to right (see arrows on the unfiltered single column image); color code: M(-polarity) = pink, N(-polarity) = blue.}
\label{fig:STEM-Fig1}
\end{figure}

The atomic scale STEM analysis of the partially switched capacitor is given in Figure \ref{fig:STEM-Fig1}. An annular bright-field (ABF)-STEM micrograph of the capacitor cross-section is depicted in Figure \ref{fig:STEM-Fig1}a. The individual thickness of the Pt electrodes is determined to be about 11 nm and 25 nm for the epitaxial Pt bottom electrode and the top electrode, respectively. They sandwich the \alscn layer with a total thickness of 8 - 9 unit cells, which was determined by counting the number of monolayers. This corresponds to 4 to 4.5 nm at a $c$-lattice parameter of $\approx 505$ pm. This number agrees well with the targeted thickness, considering the deposition rate calibrated on thicker films. Therefore, we conclude that there is no significant delay of film growth due to nucleation. The Sc content was verified by EDS analysis of a $\approx 10x4$~ nm² frame to be in the order of $x \approx 26$ at.\%.

Atomic scale investigations of the polar domain structure were conducted using ABF-STEM imaging paired with multi-frame image alignment\cite{Jones2015} on the partially switched \alscn film. Here, the use of the ABF detector allows to routinely image atomic positions of light elements such as nitrogen, which is the crucial prerequisite to observe the polarity on the unit cell level in ferroelectric \alscnXNospace.\cite{Wolff2021} Figure \ref{fig:STEM-Fig1}b depicts atomic models of the N- and M-polar oriented wurtzite-type structures sketched along the [2-1-10] viewing direction required for the investigation of unit cell polarity.
As already discussed in related work,\cite{Wolff2021,Schoenweger2022, redwan2023} sputtered nanocrystalline films of \alscnX exhibit small grain diameters of 2 - 6 nm featuring an in-plane tilt between the individual grains in the order of 6° which restricts the observable area to single grains with exact orientation to the incident electron beam.\cite{Schoenweger2022a} In this respect, the ABF-STEM image contrast formation crucially depends on exact orientation conditions.\cite{Zhou2016,Okunishi2012} In this investigation, the directly interpretable sample area was further limited by 1 - 2 nm large Pt agglomerates present evenly spaced in the center of the \alscn layer. These Pt artifacts were introduced during sample preparation using the FIB thinning method.

Individual grains with aligned zone axis orientation were identified in the \alscn layer after centering the GaN crystal lattice into the [2-1-10] orientation. The unit cell polarity was identified from the non-rigidly registered multi-frame ABF-STEM data sets, by the analyses of intensity profiles drawn across the (Al,Sc)-N dumbbells (see Supplement - Figure \ref{fig:STEM-S0} for a demonstration on the GaN substrate). The ABF-STEM micrograph contrast was inverted and a color scheme (inverted-cABF-STEM) was applied to enhance the image visibility as described in the experimental section. No noise filter was applied for the analysis of intensity profiles to avoid potential artifacts by reducing the information limit. Intensity profile analysis is regularly performed to determine the polarity of materials with wurtzite-type crystal structure because of the strong contrast difference between metal and Nitrogen or Oxygen atoms.\cite{Mata2012}

\begin{figure}[h!]
\centering
\includegraphics[]{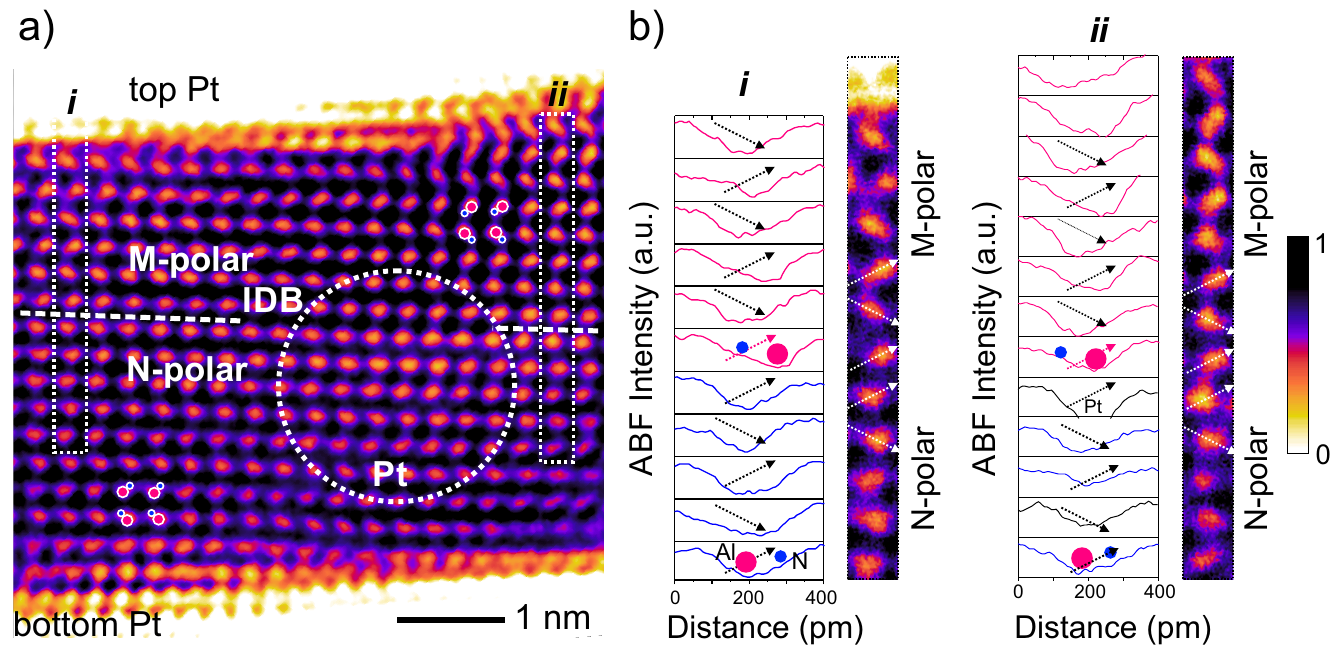}
\caption{a) Inverted-ABF-STEM micrograph of the \alscn layer featuring a horizontal inversion domain boundary separating regions of M-polarity within the top unit cells and N-polarity in the bottom film. Sketches of the (Al,Sc)-N dumbbells assist to visualize the polarity. b) Intensity profile analysis of the (Al,Sc)-N dumbbells inside the vertical single-column frames on the left ($i$) and right side ($ii$) of the grain. Both analyses hint towards the lateral progression of an IDB. Profiles are drawn from left to right (see arrows) on the unfiltered image; M(-polarity) = pink, N(-polarity) = blue.}
\label{fig:STEM-Fig2}
\end{figure}

Using the described method, the presence of N-polar and M-polar regions within a single grain is observed in Figures \ref{fig:STEM-Fig1}c and Figure \ref{fig:STEM-Fig2}. They confirm the presence of inversion domain boundaries (IDB) with a varying, yet always significant horizontal component. This is highly surprising given the fact that the horizontal component should give rise to an extreme polarization discontinuity at the domain wall, which likely requires an as-yet-unknown (charge) compensation or reconfiguration mechanism for stabilization. 
Generally, M-polarity is clearly identified in the upper region in the inverted-ABF-STEM images, while the remaining N-polarity is predominantly located at the bottom interface. Figures \ref{fig:STEM-Fig1}d and \ref{fig:STEM-Fig2}b present the aforementioned profile analysis along the highlighted vertical atomic columns showing a clear N-polar (blue profiles) polarization near the bottom interface and a switch to M-polarity (pink profiles) closer to the upper interface. At the position of polarization inversion from N- to M-polarity (the profiles are drawn in the scheme "up-down" starting left of the (Al,Sc)-N dumbbell), the alternating dumbbell orientation (and so the drawn profiles) is intercepted, hence the polarity abruptly inverses to the M-polar state following an "up-down-down-up" scheme as indicated by the arrows. This change of the polarization within the sub-5 nm grains indicates that even in very thin films with grain diameters in the single digit nm range, \alscnX gradually switches in a domain wall motion limited fashion. This suggests that the material, and possibly the wurtzite-type ferroelectrics in general, are potentially very suitable for analog switching in single-digit nanometer scaled devices. Further, as already assumed for thicker films,\cite{Wolff2021} the nucleation of polar inversion domains switching from N to M-polarity is found to be initiated at the top electrode interface and from there propagates towards the bottom interface. 
These results demonstrate the first direct observation of IDBs in wurtzite-type ferroelectrics. From the application point of view, the gradual in-grain switching and small domain size is highly attractive to address multiple states in lateral dimensions < 10 nm², which emphasizes the potential of \alscn for the realization of highly scaled synapse-emulating neuromorphic computing devices.\cite{Islam2019}

\section{Conclusion}
In summary, ferroelectric switching in sputter-deposited, 8 to 9 unit cells (4 to 4.5 nm) thin \alscn grown non-epitaxially on \substratePoly~and epitaxially on \substrateEpi~is demonstrated. The ferroelectric nature of the switching event was independently confirmed by electric $J-E$ and $C-E$ measurements, as well as by STEM investigations, resolving the polarization inversion at the atomic scale. This is the first report on a sub-5 nm thin wurtzite-type ferroelectric film switching fully on Si which also feature record low switching voltages down to $\approx 1$ V. Both aspects can be expected to greatly aid the future integration of the material class to advanced CMOS technology. Despite the better structural qualities of the thin film texture and interface structure of epitaxial films, the growth of sub-5 nm non-epitaxial \alscn on silicon results in an improved ratio between coercive- and breakdown field. Hence, the structural quality is not a limiting factor for good ferroelectric performance. 
$E_c$ in our films increased only slightly with decreasing film thickness down to 10 nm, while it decreased when the film thickness is further reduced down to sub-5 nm, thereby significantly lowering the required switching voltages. This behavior fits qualitatively to the depolarization corrected JKD model described by Dawber et al.,\cite{Dawber2003} who explain the decrease in $E_c$ by the increase in the depolarization field, resulting from the finite screening length of the electrodes. The increasing permittivity in thinner films supports this hypothesis. The permittivity is also found to increase with cycling, especially for thinner films, which we relate to an increase in the relative volume of domain walls with respect to the total film volume.

Our high resolution ABF-STEM investigation of epitaxially grown \alscn for the first time allowed to resolve IDBs in a wurtzite-type ferroelectric. The resulting presence of nanoscale domains spanning only fractions of individual nm-sized grains strongly suggests that domain wall motion still limits the switching kinetics in wurtzite-type films thinner than 5 nm. The strong horizontal component of the observed domain walls further motivates the existence of a (charge) compensation mechanism for the strong polarization discontinuity at the boundary. 

To conclude, the given evidence of in-grain switching of sub-5 nm thin films with sub-5 nm lateral grain dimensions demonstrates stable and gradual partial switching capabilities of extremely low volumes. This switching mechanism together with the positive effects of thickness downscaling on $E_c$ that result in ferroelectric switching voltages as low as 1 V make ferroelectric \alscnX a highly interesting choice for nanoscale ferroelectric synaptic devices that require analog switching - as is the usage of a CMOS BEOL compatible deposition process already used in mass-production. 

\section{Experimental section}
As electrodes, 100 nm thick Pt layers on a 10 nm thick Ti seed layer sputter-deposited on SiO$_2$/Si wafers were provided by Fraunhofer ISIT, Germany. Epitaxy-ready templates consisting of GaN(4 \textmu m)/\sapphire were commercially bought. The substrates were diced into 1x1 cm² chips with prior surface protection using a photoresist. Cleaning in acetone and isopropanol using an ultrasonic bath was performed, followed by rinsing in DI-water. Subsequently, the non-epitaxial Pt templates were cleaned by performing an Ar/O\textsubscript{2}-plasma-etching in a Sentech Sl100 reactor, details can be found elsewhere.\cite{Fichtner2017} The \alscn layers as well as the bottom epitaxial Pt and the Pt top layers were grown in-house by sputter deposition using an Oerlikon (now Evatec) MSQ 200 multisource system, details about the process can be found in a previous publication.\cite{Schoenweger2022a} The epitaxial growth on \substrateEpi~ as well as the non-epitaxial growth on \substratePoly~ was obtained by using the same deposition process. The Pt top layers were deposited $in~situ$ subsequently to the \alscn deposition after reaching a base pressure of at least 5 x $10^{-7}$ mbar. Round- as well as square top electrodes were structured with lithography and ion-beam etching (IBE, Oxford Instruments Ionfab 300). The dry-etching was stopped right after the loss of Pt signal, detected via a secondary-ion mass spectrometer (SIMS). The capacitance and loss tangent measurements were performed using a  Hewlett Packard 4284 A Precision LCR meter. If not stated otherwise, the small signal voltage and frequency were 0.1 V and 900 kHz, respectively. The sweep time for $C-E$ measurements of different \alscn thickness was kept constant by adjusting the delay time between each step, as well as the step width of the voltage sweep. $J-E$ measurements were performed using an AixACCT TF 3000 analyzer.
A cross-section sample of the partially switched film was extracted and thinned by the focused ion-beam technique using a Helios600 FIB-SEM machine and transferred into a JEOL (JEM200F) NEOARM scanning transmission electron microscope operated at 200 kV(cold-FEG). Atomic scale investigation of the unit-cell polarity within the sub-5 nm \alscn layer was conducted using the annular bright-field scanning transmission electron microscopy (ABF-STEM) mode with 10 - 20 mrad collection angle and a spatial resolution limit of $\approx 70$ pm. To minimize the effects of scan distortions and sample drift during image acquisition, the atomic-scale ABF-STEM micrographs were recorded using fast serial recording of multi-frame images followed by post-processing image alignment using rigid and non-rigid registration implemented in the Smart Align algorithm (HREM Research Inc.) on the DigitalMicrograph v.3.5.1 (DM) (GatanInc) software. If not stated otherwise, the non-registered ABF-STEM images were 1) Fourier filtered by a simple radiance difference filter using the lite version of DM plug-in HREM-Filters Pro/Lite v.4.2.1 (HREM Research Inc.) to remove high-frequency noise, and 2) the ABF contrast was inverted, a color scheme was applied and the contrast was slightly enhanced within DM, for presentation purposes (inverted-ABF-STEM). The in-plane and out-of-plane lattice parameters were estimated with ±2 pm accuracy by calculating the average atomic distance over minimum 8 unit cells and 6 unit cells, respectively, and are compared with as-determined lattice parameter of the GaN substrate. For GaN, the as-determined lattice parameters are $a \approx 318$ pm and $c \approx 521$ pm and for \alscn these are $a \approx 319$ pm and $c \approx 505$ pm. Chemical analysis on the capacitor stack was conducted using energy-dispersive spectroscopy (EDS) with a dual silicon drift detector system with 100 mm² active area each. Cross-section samples of 10 nm thin \alscn based capacitor structures grown on \substrateEpi~ and grown on \substratePoly~ were examined using a Tecnai F30 G² STwin microcsope operated at 300 kV.

\subsection{Acknowledgements}
This work was supported by the project “ForMikro-SALSA” (Grant no. 16ES1053) from the Federal Ministry of Education and Research (BMBF) and the Deutsche Forschungsgemeinschaft (DFG) under the scheme of the collaborative research centers (CRC) 1261 and 1461 as well as grant 458372836. The authors gratefully acknowledge Christin Szillus for the FIB preparation of cross-section samples for TEM analysis.

\subsection{Bibliography}
\bibliography{WileyNJD-AMA}%

\newpage
\section{Supplement}

\begin{figure}[h!]
\centering
\includegraphics[]{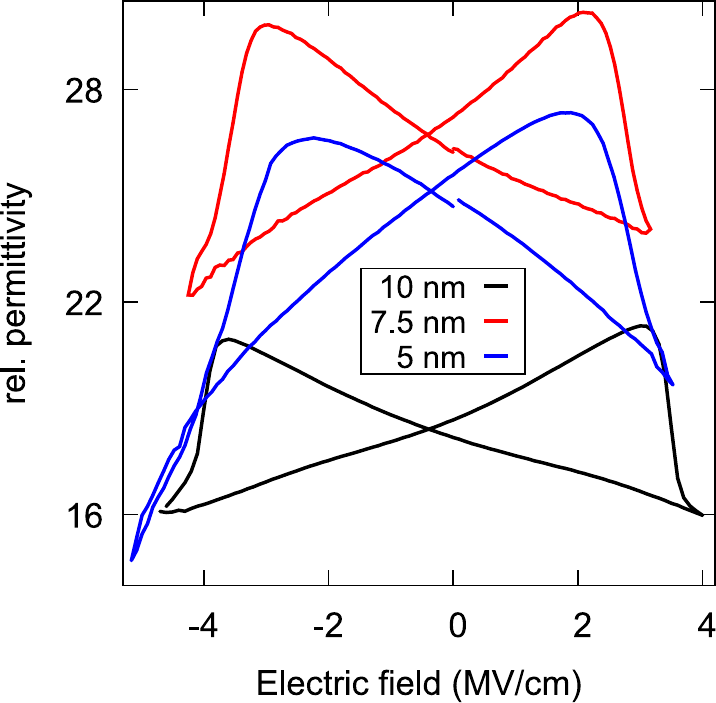}
\caption{$C-E$ loops of 10 nm - down to sub-5 nm thin layers of \alscn deposited epitaxially on \substrateEpi. The sweep time for each thickness was kept constant.  All measurements were performed on 10 \textmu m diameter pads.
No clear trend or $\varepsilon_r$ with thickness can be observed from 10 nm down to sub-5 nm film thickness due to the error in relative permittivity of $\Delta(\epsilon_r) \approx 4$, which mainly arises from variations in the capacitor area (nominal 10 µm diameter pads).}
\label{fig:C-V_GaN_thickness}
\end{figure}

\begin{figure}[h!]
\centering
\includegraphics[]{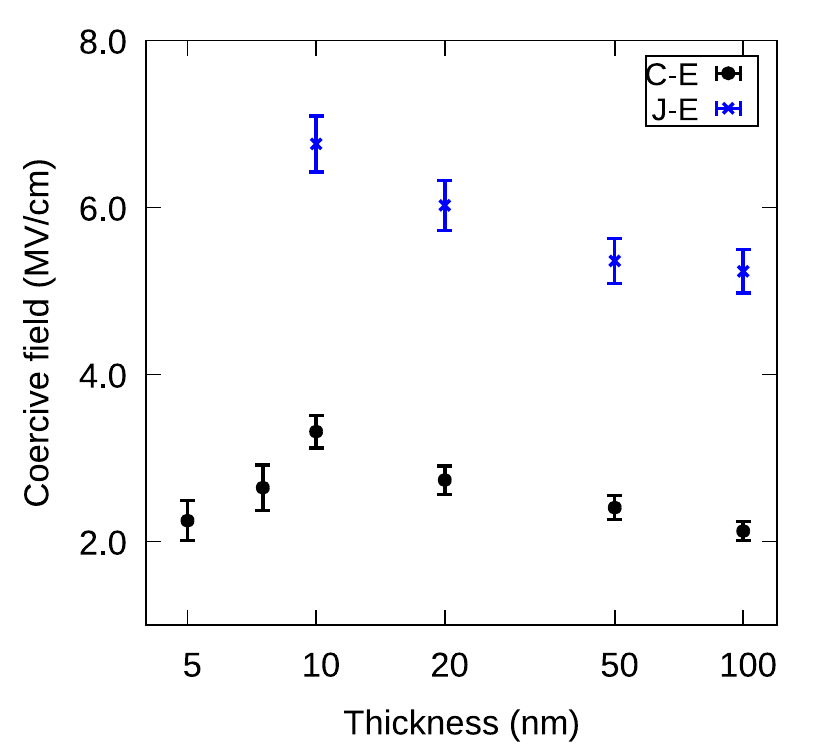}
\caption{The coercive field of \alscn epitaxially grown on \substrateEpi~ in dependence on film thickness. $E_c$ is determined via $J-E$ (80 kHz) and via $C-E$ (sweep time 20 s, small signal 100 mV and 900 kHz) loops. The breakdown field approaches the coercive field below 10 nm film thickness at high frequencies, thus not allowing to clearly determine $E_c$ below that film thickness.}
\label{fig:Supp.Ec-vs-thickness-epiPt}
\end{figure}

\begin{figure}[h!]
\centering
\includegraphics[]{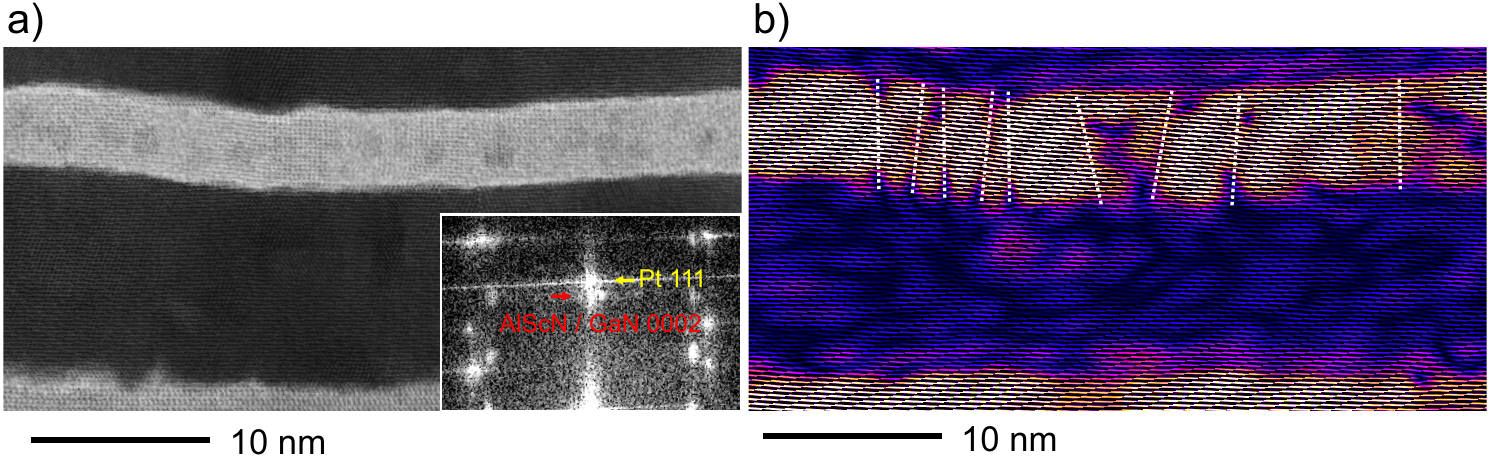}
\caption{Nanoscale texture analysis of single-digit nanometer large columnar grains. a) ABF-STEM micrograph of the film cross section with Fast Fourier Transform (FFT) showing the aligned out-of-plane Pt 111 and \alscn 0002 reflections. b) Inverse FFT image using an aperture to filter the spatial frequencies of the GaN and \alscn out-of-plane 0002 reflections to demonstrate the crystal size and c-axis alignment. Dotted lines indicate the position of grain boundaries.}
\label{fig:FigS_texture}
\end{figure}

\begin{figure}[h!]
\centering
\includegraphics[]{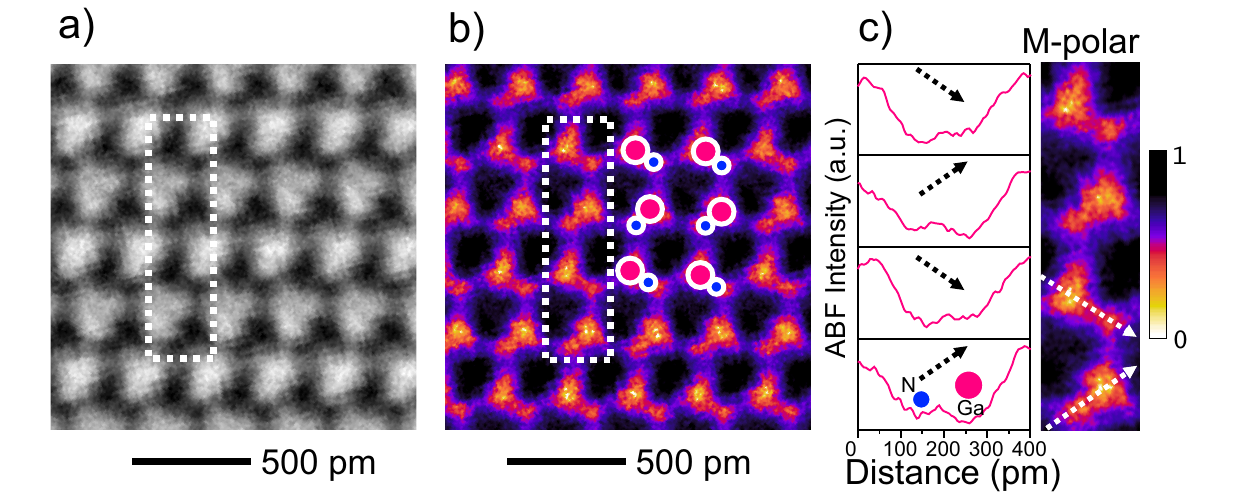}
\caption{a) Unfiltered ABF-STEM micrograph of the GaN substrate and b) Inverted-ABF-STEM image featuring M-polarity. Sketches of the Ga-N dumbbells assist to visualize the M-polarity. c) Display of intensity profile analysis of the Ga-N dumbbells inside the vertical frame. Profiles are drawn from left to right (see arrows) on the unfiltered image; M(-polarity) = pink, N(-polarity) = blue. }
\label{fig:STEM-S0}
\end{figure}

\begin{figure}[h!]
\centering
\includegraphics[]{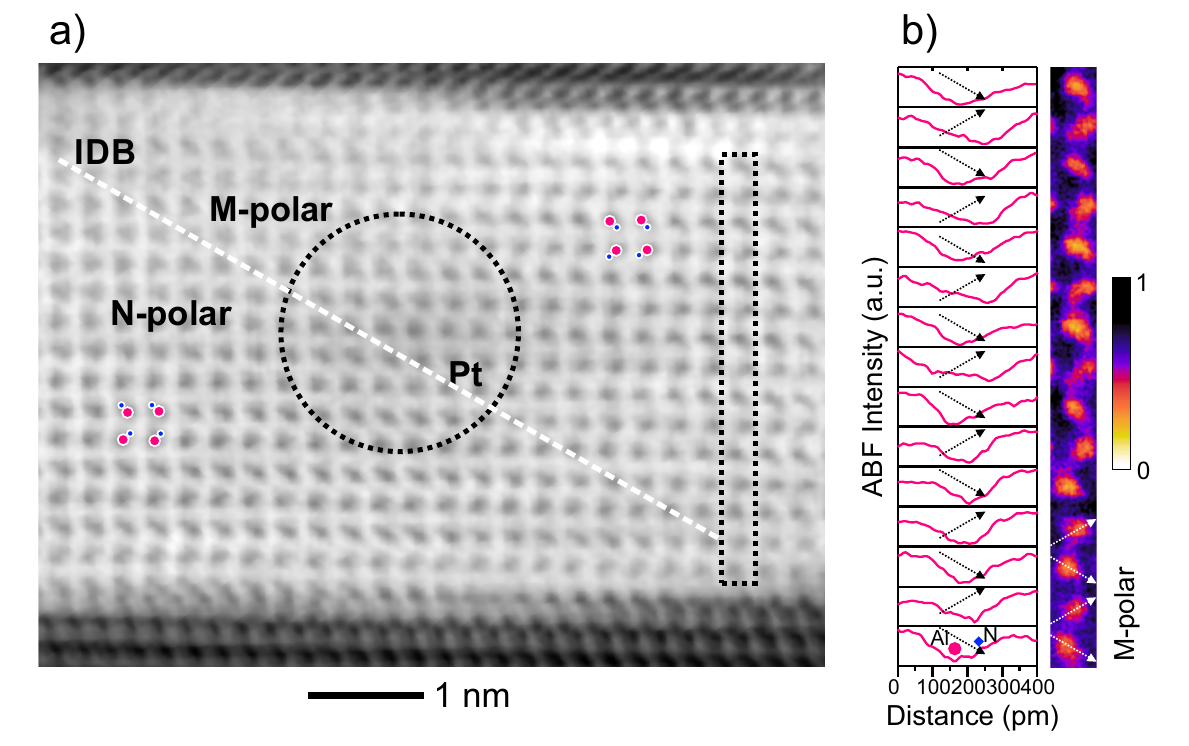}
\caption{a) Unfiltered grayscale ABF-STEM micrograph of the inverted-ABF-STEM image shown in Figure \ref{fig:STEM-Fig1} including the sketched position of the inclined inversion domain boundary. b) Profile analysis from the vertical single column frame where the M-polar domain reaches down to the bottom Pt interface. Profiles are drawn from left to right (see arrows) on the unfiltered image; M(-polarity) = pink, N(-polarity) = blue. }
\label{fig:STEM-S1}
\end{figure}

\end{document}